\documentclass{appolb}
\usepackage{epsf}
\usepackage{epsfig}
\usepackage{graphicx} 
\usepackage{amsmath}
\usepackage{amsfonts,amsbsy}
\usepackage{amssymb}
\usepackage{graphicx}
\usepackage{bm}
\usepackage{resizegather}
\usepackage[colorlinks=true,pdfstartview=FitV,linkcolor=red,citecolor=blue,urlcolor=blue]{hyperref}

\def\be{\begin{equation}}
\def\ee{\end{equation}}
\def\bea{\begin{eqnarray}}
\def\eea{\end{eqnarray}}

\mathchardef\Re="023C
\mathchardef\Im="023D

\begin{document}

\title{ From the Glasma to the QCD Phase Boundary
\thanks{Invited talk presented at Critical Point and Onset of De-Confinement 2016 held in Wroclaw, Poland, May 30-June 4, 2016}
}
\author{Larry McLerran
\address{Institute for Nuclear Theory, University of Washington, Box 351550, Seattle, WA, 98195, USA}
\address{China Central Normal University, Wuhan, China}
}
\maketitle

\begin{abstract}
In this paper, I qualitatively discuss the matter formed in the fragmentation region of nuclear collisions at the highest energies.  I argue that although the initial temperature and baryon number density can become very large, the ratio of initial baryon chemical potential to initial temperature,
$\mu_B/T$ is approximately independent of energy, when measured at a fixed rapidity measured from the end of the fragmentation region..  This quantity is argued to be roughly invariant under expansion,
and therefore the value measured at decoupling should be approximately the same as the initial value and largely independent of energy. The values of the initial temperature and initial baryon number are energy dependent and become  large as the center of mass collision energy increases.. 
\end{abstract}

\maketitle

\section{Introduction}

The title of the talk was given to me by the organizers and force me to rethink what implications gluons saturation\cite{McLerran:1993ni}-\cite{McLerran:1993ka} and the Glasma\cite{Kovner:1995ja}-\cite{Kovner:1995ts} might have to do with the phase diagram of QCD in the region of high baryon number density.  The problem I will address is what values of baryon number density and temperature are probed in  the fragmentation region of nucleus-nucleus collisions at asymptotically high energy.  I was therefore forced to go back and update the ideas found in the paper by Anishetty, Koehler and McLerran\cite{Anishetty:1980zp}, and put hem in a more modern context.  This work is largely based on and  stimulated by  the considerations of Li and Kapusta\cite{Li:2016wzh}.
This region was studied emperically based on experimental data in the classic work of Cleymans and Becattini\cite{Becattini:2007qr}.
I have attempted to simplify some of the arguments in these very nice papers, and address specifically the region of the phase diagram of QCD that one can study at asymptotic energies.

The result I find by very simple kinematic arguments is amusing:  At asymptotically high energies and fixed rapidity as measured from the end of the fragmentation region, one makes a system initially at a very high temperature and a very high baryon number density.  This temperature and baryon density grow as the collision energy increases.
This can be much higher than the expected scale of the de-confinement or chiral symmetry restoration temperatures and densities.  The ratio of the baryon chemical potential to temperature asymptotes to a constant.  Since baryon number is conserved and the entropy is approximately conserved during expansion, the expansion dynamics should little change this ratio.  Therefore at late stages of the collision, one is probing the baryon rich region of the phase diagram of QCD.

Although the experimental challenge to probe such a region is formidable, the theoretical advantage of studying such a region is that at a fixed temperature and baryon number density, the matter produced at high energy is more slowly expanding than at lower energy.  This gives longer time for long range correlations to grow.  In addition, I believe we understand the dynamics of the early times better at high energy than we do at low energy.

\section{Some Phenomenological Considerations}

\begin{figure}
	\centerline{\includegraphics[width=0.5\linewidth]{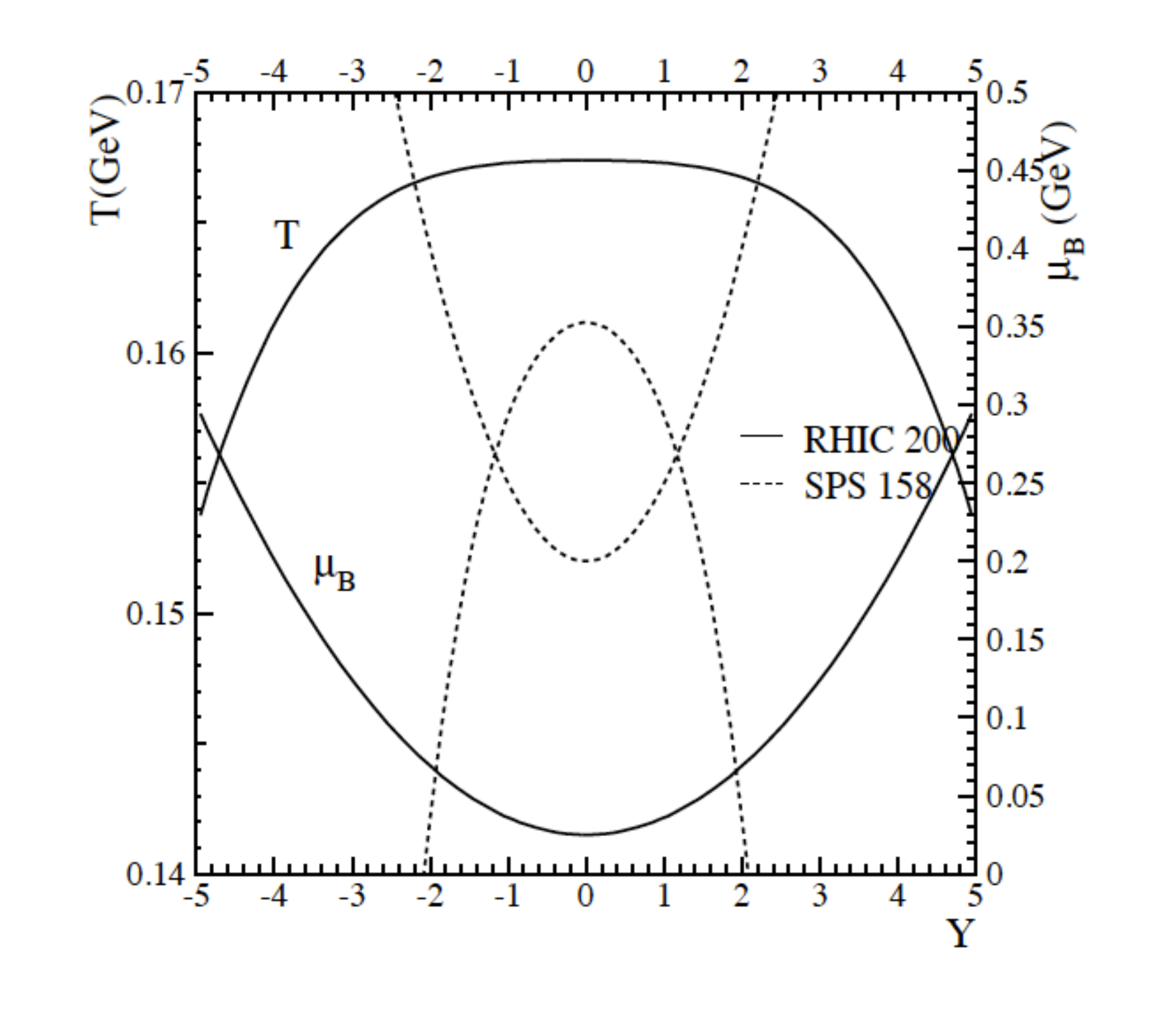}}
\caption{Values of temperature and  baryon number chemical potential as a function of rapidity for SPS and RHIC energies, from Ref.  \cite{Becattini:2007qr}   }
\label{cleymans}
\end{figure}

In Fig. \ref{cleymans},  the values of baryon chemical potential and temperature extracted from fits to the spectra of produced particles at the top SPS and RHIC energies are shown.  The ratio of $\mu_B/T$ for systems where the baryon number density is small compared to the temperatures is up to a constant the ratio of baryon number density to entropy.
\begin{equation}
   \mu_B/T \sim N_B/S
\end{equation}
Since baryon number is conserved during expansion and entropy is approximately conserved if the effects of viscosity are ignored, this ratio is an approximate invariant under expansion.  If the system has initially a very high temperature then the initial value of $\mu_B$ must be correspondingly increased.  The advantage of studying systems that are initially very hot and dense is that by the time they reach low values of density and temperature, they are expanding more slowly than systems that started at lower temperatures and densities.  This is because characteristic volumes and times are larger at fixed energy density for systems with higher initial densities.

If there was limiting fragmentation for both baryons and pions, then we would expect that the rapidity distributions of these particles would be energy independent when measured as a function of the rapidity
distance from the kinematic limit for nucleon-nucleon scattering.  Such limiting fragmentation has been observed at RHIC energies.  

The question we will address is how does the initial energy density and baryon number density scale with energy, and how does this independence of $\mu_B/T$ arise.

\section{Using Saturation to Estimate the Early Baryon Density and Temperatures}

In the CGC description of the initial condition for  ultra-relativistic heavy ion collisions,
the saturation momentum grows as a function of the rapidity distance from the fragmentation region as an exponential in rapidity (power law in 1/x).  If we sit in the fragmentation region of one heavy ion at some fixed rapidity difference from the beam rapidity, the density of partons in the beam nucleus does not grow.  Let us call this rapidity difference $\Delta y$.  The saturation momentum of the first nucleus is fixed at
\begin{equation}
    Q_1^2 \sim Q_0^2 e^{\kappa \Delta y}
\end{equation}
where phenomenologically $\kappa$ is a number of order $\kappa \sim 0.2-0.3$.
The saturation momentum in the other nucleus grows as the beam energy like
\begin{equation}
   Q_2^2 \sim Q_0^2 e^{\kappa(2y_{cm} - \Delta y)}
\end{equation}
where $y_{cm}$ is the center of mass rapidity.

At the saturation momentum of the first nucleus, the second nucleus is completely saturated and appears as a black disk.  The collision will strip all of the partons from nucleus one up to a typical momentum scale of $Q_2^{sat}$.  In so far as the parton distribution have little dependence upon
the momentum scale of their measurement, the multiplicity will only very weakly depend upon $Q_2^{sat}$ and therefore the beam energy.  There is therefore approximate limiting fragmentation.

Let us estimate the density of produced particles in the fragmentation region of nucleus one.
For  central collisiion of two nuclei of size $R$,  multiplicity per unit area scales as
\begin{equation}
  {1 \over {\pi R^2}} {{dN} \over {dy}} \sim Q_1^2
\end{equation}
The initial time and initial longitudinal length scales as $Q_2^{-1}$, so that the initial entropy density is 
\begin{equation}
       {S_{initial} \over V} \sim Q_1^2 Q_2
\end{equation}

\begin{figure}
	\centerline{\includegraphics[width=0.5\linewidth]{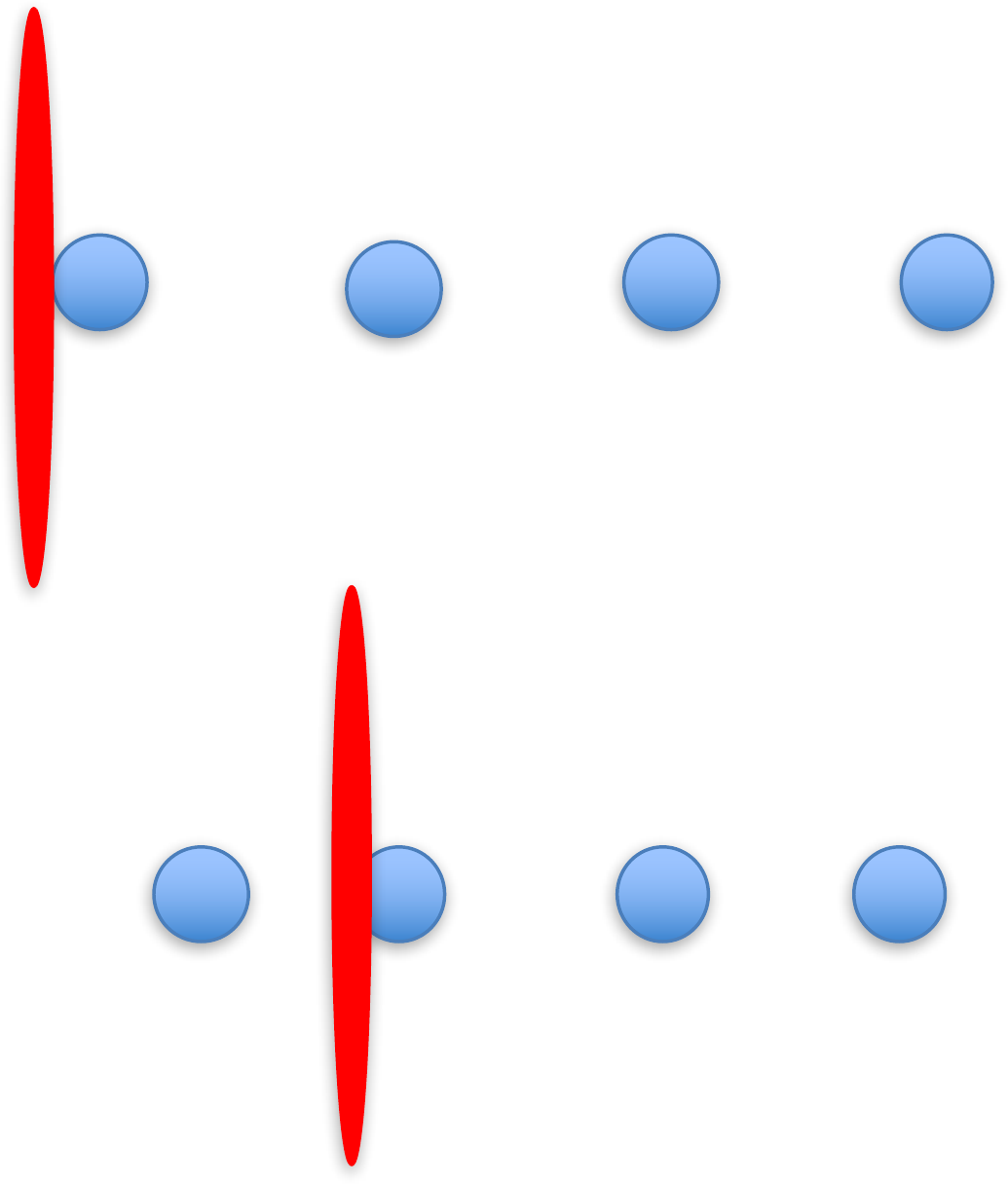}}
\caption{The projectile nucleus passing through a target at rest  }
\label{sequential}
\end{figure}

The initial baryon density is caused by compression of the target nucleon (nucleus 1) as the projectile (nucleus 2) passes through the target.  This is easiest seen in the target rest frame.  There the projectile is a thin disk striking a row of target nucleons, Fig. \ref{sequential}.
The compression appears to be $1/(1-v)$ where $v$ is the velocity of the projectile, but the struck nucleons are moving and the true compression is computed in the rest frame, which reduced the compression by a factor of $\gamma_{comp}^{-1}$, resulting in a real compression factor of $\gamma_{comp}$

Now we need to determine the velocity associated with this compression.  This is also simple to estimate since the longitudinal energy of fragments is trapped inside the nuclear fragmentation region so long as it is formed inside the target.  This means that
\begin{equation}
\gamma_{comp} \sim Q_2 R_{1}
\end{equation}
since the typical transverse mass scale for a particle is of the order of the saturation momentum of the projectile nucleus. We have therefore that
\begin{equation}
   {{N_B} \over V} \sim Q_2 R_1
\end{equation}
or that  $N_B/S$ is independent of the saturation momentum of nucleus 2 and is proportional to $R_1/Q_1^2$  This ratio  has dependence on the rapidity difference from the kinematic endpoint, $\Delta y$ ,but is independent of $R_1$.
These arguments are of course too crude to accurately resolve the rapidity dependence of this ratio.  To do this properly requires a detailed simulation of the particle production and baryon compression factors.

\section{Summary and Conclusions}

The simple conclusion from this analysis is that as one makes the collision energy asymptotically higher,
then at some fixed distance away from the kinematic end point associated with the energy per nucleon in a heavy ion collisions, the produced matter is initially hotter, denser and produced earlier.
The entropy per baryon is roughly independent of energy, so that at late times at fixed energy density
the matter being studied is independent of beam energy.  However the dynamics of expansion will have changed and the matter is expanding more slowly as the energy changes.  Therefore from  a theorists perspective, studying matter at finite baryon density is perhaps simpler as the energy is increased.  Unfortunately, it gets much more almost diffiicult for experimentalists

\section{Acknowledgements}
L. McLerran was  supported under Department of Energy contract number Contract No. DE-SC0012704 at Brookhaven National Laboratory, and L. McLerran is now supported under
Department of Energy under grant number DOE grant No. DE-
FG02-00ER41132.

\end{document}